\documentclass[twocolumn,preprintnumbers,amsmath,amssymb,aps,prb,longbibliography]{revtex4-1}
\usepackage{graphicx}
\begin{document}

\title{Commensuration Effects on Skyrmion Hall Angle and Drag for Manipulation of Skyrmions on Two-Dimensional Periodic Substrates  }
 
\author{C. Reichhardt and C. J. O. Reichhardt}
\affiliation
{
Theoretical Division and Center for Nonlinear Studies,
Los Alamos National Laboratory, Los Alamos, New Mexico 87545, USA}

\date{\today}

\begin{abstract}
We examine the dynamics of an individually driven skyrmion moving through 
a background lattice of skyrmions coupled
to a 2D periodic substrate
as we vary
the ratio of the number of skyrmions to the number of pinning sites
across commensurate and incommensurate 
conditions.
As the skyrmion density increases, the skyrmion Hall angle
is nonmonotonic, dropping to low or zero values in commensurate states and
rising to an enhanced value in incommensurate states.
Under commensuration, the driven skyrmion
is channeled by a symmetry
direction of the pinning array and exhibits an increased velocity.
At fillings for which the skyrmion Hall angle is zero,
the velocity has a narrow band noise signature, while for incommensurate
fillings, the skyrmion motion is disordered and the velocity noise is broad band.
Under commensurate conditions, multi-step depinning transitions appear
and the skyrmion Hall angle is zero at low drives
but becomes finite at higher drives, while at incommensurate
fillings there is only a single depinning transition.
As the gyrotropic component of the skyrmion dynamics, called the Magnus force,
increases,
peaks in the velocity that appear
in commensurate regimes cross over to  dips,
and
new types of directional 
locking effects can arise
in which the skyrmion travels along other symmetry directions of the
background lattice.
At large Magnus forces,
and particularly at commensurate fillings,
the driven skyrmion can experience a velocity boost
in which
the skyrmion moves faster than
the applied drive due
to the alignment of the Magnus-induced velocity
with the driving direction.
In some cases, an increase of the
Magnus force can produce regimes of enhanced pinning when the skyrmion 
is forced to move along a non-symmetry direction
of the periodic pinning array.
This is in contrast to  systems 
with random pinning,
where increasing the Magnus force generally reduces the pinning effect.
We demonstrate these dynamics for both square and triangular substrates,
and map out the different regimes as a function of filling fraction,
pinning force, and the strength of
the Magnus force in a series of dynamic phase diagrams.

\end{abstract}

\maketitle

\section{Introduction}

Magnetic skyrmions are particle-like spin textures found in numerous systems
\cite{Muhlbauer09,Yu10,Nagaosa13,Fert17,Jiang17a,EverschorSitte18,Birch20,Bogdanov20}, including materials in which
the skyrmions are stable at
room temperature \cite{Jiang15,MoreauLuchaire16,Boulle16,Soumyanarayanan17}.
Skyrmions can also be generated by an applied current
\cite{Jonietz10, Zang11, Yu12, Lin13a, Liang15, Woo16,Legrand17,Montoya18}.
Due to their size scale, mobility, and stability, skyrmions
are promising candidates for a variety of
applications \cite{Fert13,Tomasello14,Pinna20,Song20}, many of which 
require the ability to control how
the skyrmions move, how they interact with defects or nanostructures,
and how to manipulate them on the individual level.
The skyrmions also interact with
quenched disorder in the system,
giving rise to a pinning effect and establishing
a threshold driving force that must be applied for the skyrmion to be set in
motion \cite{Reichhardt21, Liu13, Iwasaki13,Salimath19,Xiong19,Juge19,DelValle22}.
The magnitude of this
threshold depends on the properties of the disorder
\cite{Reichhardt21,Stosic17,Fernandes18,Navau18}, collective interactions
with other skyrmions \cite{Koshibae18},
and thermal effects \cite{Lin13,Reichhardt18a,Litzius20}.
There are also
other types of skyrmions and skyrmion-like textures \cite{Gobel21a},
including ferromagnetic \cite{Muhlbauer09,Yu10} and
antiferromagnetic \cite{Barker16, Legrand20} skyrmions,
antiskyrmions \cite{Nayak17,Kovalev18},
and merons \cite{Yu18a}.

For ferromagnetic skyrmions
subjected to an
applied drive,
the skyrmion motion 
exhibits a skyrmion Hall effect along the skyrmion Hall angle \cite{Nagaosa13,EverschorSitte14,Brearton21} $\theta_{sk}$, which arises
from the topology of the skyrmions.
The intrinsic skyrmion Hall angle $\theta^{\rm int}_{sk}$ is proportional to the ratio of the
Magnus or gyrotropic term to the dissipative term of the skyrmion dynamics.
Many proposed skyrmion applications
require the reduction or absence of the skyrmion Hall effect, 
so there have been numerous studies
focused on understanding
how to control the skyrmion Hall angle, such as
by the use of nanostructures 
\cite{Reichhardt21,DelValle22}. The magnitude of the skyrmion Hall effect
is modified both by the pinning landscape and by the velocity of the skyrmions.
Quenched disorder induces a
side jump effect
that reduces the skyrmion Hall angle below its intrinsic value.
The side jump is largest
for small velocities just above depinning,
giving a skyrmion Hall angle that
is zero at depinning and increases with increasing drive
until it saturates at a value near $\theta^{\rm int}_{sk}$
for high drives
\cite{Muller15, Reichhardt15a, Reichhardt15,Jiang17, Legrand17,Kim17, Diaz17, Juge19,Litzius20, Fernandes20,Yu20,Peng21}. 
The skyrmion Hall angle can also be affected
by shape distortions of the skyrmions and collisions with other skyrmions
\cite{Litzius17,Litzius20, Juge19, Zeissler20}.
In addition to the importance of understanding skyrmion dynamics in the
presence of disorder for applications,
the strong gyrotropic
nature of the dynamics means that
skyrmions represent a new class of systems that can show collective dynamics when 
driven over random or periodic substrates.
Most previous studies of such behavior have involved
overdamped systems \cite{Reichhardt21}. 

One approach to generating well-controlled skyrmion motion is
to couple the skyrmions to nanostructures, such as a
periodic array of defects
or other types of engineered landscapes
\cite{Fert13,Tomasello14,Reichhardt21,Fernandes20a,Migita20,Gobel21,Juge21}.
In this case, it is important to understand how individual skyrmions
interact with both the defect array and
with the other skyrmions.
An example of such a system is 
skyrmions interacting  
with a two-dimensional (2D) periodic substrate,
where the system can be characterized by a filling factor $f$
corresponding to the ratio of the number of skyrmions $N_{s}$
to the number of pinning sites $N_{p}$.
When $f = N_{s}/N_{p}$ is an integer,
the skyrmions form a commensurate or 
ordered crystalline structure.
Commensuration effects for particles on 2D
substrates have been studied extensively in other condensed 
matter systems, such as the ordering of atoms or molecules
on surfaces \cite{Bak82}, 
sliding friction \cite{Vanossi13}, colloidal particles  
on patterned substrates \cite{Reichhardt02a, Mangold03},
dusty plasmas  \cite{Huang22}, Wigner crystal ordering
in moir{\` e} systems \cite{Xu20},
vortices in Bose Einstein condensates \cite{Tung06}, and cold atoms 
coupled to  optical traps \cite{Lewenstein07}.
The closest match to the skyrmion system, however, is
vortices in type-II superconductors coupled to 2D pinning arrays
\cite{Baert95,Harada96,Martin97,Reichhardt98a,Berdiyorov06,Sadovskyy17}. 
At commensuration,
the superconducting vortices
form an ordered lattice and exhibit a strong enhancement 
of the pinning effect observable as
peaks in the depinning threshold as a function of changing superconducting
vortex density. 

Commensurate-incommensurate systems display a rich variety of dynamical phases
since the collective motion differs at commensurate and incommensurate
fillings
\cite{Reichhardt21}.
For example, there can be multiple step depinning,
transitions between
ordered and disordered flow
\cite{Gutierrez09,Vanossi12}, and soliton motion \cite{Bohlein12}.
Particle flow on 2D  periodic substrates
can be modified significantly depending on the direction of drive with respect to
symmetry directions of the underlying substrate.
For example,
in directional or symmetry locking,
the particles preferentially
move along certain symmetry directions of the
pinning lattice even when the drive is not aligned
with those directions,
and as a result, for changing drive orientation a series of
steps appear in the velocity versus driving angle curves
\cite{Reichhardt99,Korda02,Silhanek03,Gopinathan04,Cao19,Stoop20}.
Since the skyrmion Hall angle depends on the magnitude of the drive in
systems with pinning,
when individual skyrmions move
over 2D periodic substrates,
numerical studies have shown
that the skyrmion motion locks to different substrate symmetry directions
as the magnitude of the drive increases
\cite{Reichhardt15a,Vizarim20,Feilhauer20}.
For a square array, such locking directions include
$0^\circ$ and $45^\circ$ from a primary lattice vector.
In general, locking of skyrmions moving on a square array
can occur at angles $\phi = \arctan(n/m)$ from the
primary symmetry axis, with integer $n$ and $m$;
however,
the size of the pinning sites as well as interactions with other skyrmions
can limit which symmetry directions are accessible.
Other studies for skyrmions on 2D pinning arrays
indicate that different types of crystalline ordering occur at the
matching fields \cite{Ma16} and
that large scale collective flow states can arise under bulk driving \cite{Reichhardt18}.
In micromagnetic simulations, skyrmions moving on a 2D pinning array 
exhibit a number of different dynamic phases that
could be useful for applications \cite{Zhang21a}.
In addition, there are now various experiments on
skyrmion states in periodic one-dimensional (1D)  \cite{Juge21}
and 2D pinning arrays \cite{Saha19}.

In this work, we examine the dynamics of skyrmions
on a 2D square or triangular pinning lattice.
We drive a single skyrmion that interacts both
with the other skyrmions and directly with the substrate,
and measure
the velocity and direction of motion of the driven skyrmion 
as the system passes through
a series of commensurate-incommensurate transitions
at varied pinning strength and
varied ratios of the Magnus term to the dissipative term.
This work builds upon our previous studies examining the dynamics of individually 
driven
superconducting vortices \cite{Ma18,Reichhardt21b} and
skyrmions \cite{Reichhardt21a,Reichhardt21c}
interacting with either a background lattice of particles or with pinning.
In the case of superconducting vortices
where the motion is overdamped, numerous methods to drive 
individual vortices, including nanotips \cite{Straver08,Auslaender09}
and optical trapping \cite{Veshchunov16,Kremen16}, 
have been studied in experiments and simulations.
Individual skyrmions can also be driven 
with different types of tips \cite{Hanneken16},
local magnetic field gradients \cite{Wang17,Casiraghi19}, 
and with optical trapping
\cite{Yang18,Wang20b,Hirosawa22}.
The method of driving
individual particles though a background of
other particles while measuring the drag on the driven particle from
fluctuations
is known as active rheology and has been studied experimentally and 
theoretically for
colloidal particles
\cite{Habdas04,Squires05,Dullens11,Gazuz09, Zia18},
granular matter \cite{Drocco05,Candelier10,Kolb13},  
active matter \cite{Reichhardt15b},
and superconducting vortex systems
\cite{Reichhardt08,Reichhardt09a,Auslaender09,Ma18}. 

In most active rheology studies, under a
constant driving force the velocity of the 
driven particle decreases as the density of the system increases
due to an increase in the frequency of collisions with background
particles,
and there can be a sudden drop
to zero motion or a pinning transition when the system passes through
a critical density into a glass, jammed, crystalline, or amorphous solid
state.
In our previous work on active rheology
in a skyrmion system,
we considered a single skyrmion driven through
a background of other skyrmions
in the absence of pinning \cite{Reichhardt21a}.
For a constant driving force, we found that
the skyrmion Hall angle decreases with increasing
skyrmion density due to enhanced collisions; however,
particularly for systems with a strong Magnus force,
we also found a counter-intuitive increase in the velocity,
or a boost effect, in which
the skyrmion velocity increases with increasing system density. 
In some cases, the skyrmion velocity is larger than what it would be
in the absence of collisions with other skyrmions. 
This boost effect arises from
a combination of the skyrmion Hall effect and 
density fluctuations created in the surrounding
skyrmions by the driven skyrmion.
The density gradient
forms perpendicular to the direction of the drive and exerts a repelling
force on the driven particle along this direction,
but the Magnus term generates a velocity perpendicular to this repelling
force and parallel to the drive.
This is example of what 
is known as an odd-viscosity effect of the type observed in chiral
systems with gyroscopic forces
\cite{Reichhardt22,Banerjee17,Soni19,Reichhardt19a}. 
We have also considered single driven skyrmions interacting with
other skyrmions in the presence of random quenched disorder,
where in addition to velocity boost phenomena, we observe
several pinned and jammed phases as well as
stick-slip motion \cite{Reichhardt21c}.

For overdamped systems we have also numerically examined 
the active rheology of superconducting vortices and
colloidal particles interacting with 2D periodic pinning arrays 
as the filling factor of the system is varied. Here we observe what we call
an anti-commensuration effect in which
the drag on the driven particle is reduced at commensurate matching 
conditions \cite{Reichhardt21b}, opposite from the behavior found
in bulk driven systems
\cite{Baert95,Martin97,Reichhardt98a,Berdiyorov06,Sadovskyy17}. 
The drag reduction at commensuration appears
when the surrounding particles become strongly coupled to the substrate at
matching conditions and cannot be dragged along by the driven particle,
whereas at incommensurate fillings, the surrounding
particles are much more weakly coupled to the substrate, permitting
the driven particle to drag background particles and increasing
the effective viscosity it experiences. 
The result is a strongly non-monotonic drag
that shows a series of peaks at the matching conditions.
In the case of an individual skyrmion driven over a 2D periodic array
at commensurate and 
incommensurate conditions, the Magnus force
produces much more complex dynamics than are found in
the overdamped superconducting vortex system \cite{Reichhardt21b}. 

In this work we demonstrate that active rheology
for a skyrmion driven through a background lattice in the
presence of a 2D periodic pinning array 
produces very different behavior from that found
for random pinning \cite{Reichhardt21c}
or in the absence of pinning \cite{Reichhardt21a}.
The skyrmion Hall angle is non-monotonic,
falling to zero at commensurate conditions
where the skyrmions form an ordered lattice and the skyrmion
velocity peaks.
For strong Magnus forces,
the skyrmion Hall angle remains finite but is still reduced
at the matching conditions.
In general, the skyrmion motion is ordered at commensurate conditions 
and disordered at incommensurate fillings. 
For increasing Magnus force,
there are more extended regions of finite drive
for which the skyrmion
remains pinned, a behavior that is
the opposite of what is found for bulk driven systems with random pinning. 
At strong Magnus forces,
under commensurate conditions
we observe a 
pronounced velocity
boost effect
where the driven skyrmion
moves faster than it would if there were no substrate and no
collisions with other skyrmions.
This  
boost occurs when the
direction of motion of the driven skyrmion
becomes locked
along an interstitial channel and
the driven skyrmion experiences
a perpendicular repulsion from the skyrmions trapped in the pinning sites,
which is converted by the Magnus term to a velocity
in the direction of drive.
This effect is similar to the velocity enhancement found
for skyrmions moving along sample edges \cite{Iwasaki14,CastellQueralt19}. 
For higher Magnus forces,
the motion becomes increasingly chaotic
and the effect of the substrate is strongly reduced.

\section{Simulation and System}
We consider a 2D system of size $L \times L$
with periodic boundary conditions in the $x$ and $y$ directions
containing a square pinning array with lattice constant $a$.
The total number of pinning sites is $N_{p}$,
giving a pinning density of $n_{p} = N_{p}/L^2$.
The sample contains  $N_{s}$ interacting skyrmions that are modeled
as point particles according to a
modified Thiele equation \cite{Lin13,Reichhardt15}, in which the
skyrmions have repulsive interactions with each other and
attractive interactions with the pinning sites.
We characterize the system by a filling factor $f=N_{s}/N_{p}$. 
For integer values of $f$, the
system is commensurate and adopts a defect-free crystalline ordering,
while for incommensurate fillings, the system either
becomes amorphous or
forms a crystalline state containing
interstitials or vacancies \cite{Reichhardt21}.
For certain fractional fillings, such as $f = 1/2$
or $f=3/2$, the system
can be partially ordered \cite{Grigorenko03,Reichhardt01a}.
Each pinning site has a finite spatial extent, so
individual skyrmions can sit either inside a pinning site or in the
interstitial regions between the pinning sites
depending on the filling factor and the pinning strength.
From previous studies of superconducting vortices interacting with
2D square periodic pinning arrays, it is known that the particles will form a square 
lattice with all of the pinning sites occupied for $f = 1.0$,
a checkerboard pattern with half of the particles in interstitial sites
at $f = 2.0$,
an ordered lattice of dimers at $f = 3.0$,
and a hexagonal lattice at $f = 4.0$
\cite{Harada96,Reichhardt98a,Berdiyorov06,Duzgun20}.
The initial skyrmion positions are obtained by performing simulated annealing
from a high temperature molten state
down to $T = 0$, as in previous work \cite{Reichhardt98a}.
After the system is initialized,
we insert an additional interstitial skyrmion that is
coupled to an applied driving force. This
driven particle interacts both with the pinning sites
and with the other skyrmions.

The equation of motion for skyrmion $i$ is given by
\begin{equation}
	\alpha_d {\bf v}_i  + \alpha_{m}{\hat {\bf z}} \times {\bf v}_i =  {\bf F}^{ss}_i + {\bf F}_i^{p} + {\bf F}^{D}_i
\end{equation}
where the skyrmion velocity is ${\bf v}_i=d{\bf r}_i/dt$.
The first term on the left with damping constant $\alpha_d$ is the dissipation
that aligns the skyrmion motion in the direction of
the net applied force.
The second term on the left
is the Magnus force
of magnitude $\alpha_{m}$
that generates a velocity component perpendicular
to the net force.
The skyrmion-skyrmion interaction is described by
${\bf F}^{ss}_{i} = \sum_{j = 1}^{N_s} K_1(r_{ij}) \hat{\bf{r}}_{ij}$,
where the distance between skyrmion $i$ and skyrmion $j$
is $r_{ij} = |{\bf r}_i - {\bf r}_j|$,
$\hat{{\bf r}}_{ij} = ({\bf r}_i - {\bf r}_j)/r_{ij} $,
and $K_1$ is the modified
Bessel function which decays exponentially for large $r$ \cite{Lin13}.
The pinning sites
are modeled as finite-range
parabolic potential traps of radius $r_p$
that exert a maximum pinning force of $F_{p}$,
giving
${\bf F}_i^{p} = \sum_{k = 1}^{N_p} (F_p/r_p) ({\bf r}_i - {\bf r}_k^{(p)}) \Theta( r_p - |{\bf r}_i - {\bf r}_k^{(p)}| ){\hat {\bf r}_{ik}^{(p)}}$,
where  $\Theta$ is the Heaviside step  function.
We fix the pinning density to $n_p=N_p/L^2=0.4882$ throughout this work.
The driving force is ${\bf F}^{D} = F_{D} \hat{\bf{x}}$ for the driven skyrmion
and ${\bf F}^D=0$ for all of the other background skyrmions, and the driving
is always applied along the positive $x$ direction.
In the absence of other skyrmions or pinning, the 
driven skyrmion will move
with an intrinsic skyrmion Hall angle
of $\theta^{\rm int}_{sk} = \arctan(\alpha_{m}/\alpha_{d})$.
Pinning and skyrmion-skyrmion collisions can
modify the observed skyrmion Hall angle,
$\theta_{sk} = \arctan(\langle V_{y}\rangle/\langle V_{x}\rangle)$,
where $\langle V_{x}\rangle$ is the average velocity along
the driving direction and
$\langle V_{y}\rangle$ is the average velocity perpendicular to the drive.
For convenience, we use the
normalization condition $(\alpha^2_{d} + \alpha^2_{m})^{1/2} = 1.0$.
This constraint ensures
that
we always have
$\langle V\rangle = (\langle V_x\rangle^2 + \langle V_y\rangle^2)^{1/2} = 1.0$
for $F_{D} = 1.0$ in the absence of pinning
regardless of the value of $\theta_{sk}^{\rm int}$.
It also makes a velocity boost easy to detect,
since for example if $F_{D} = 1.0$, there is a boost
whenever $\langle V\rangle > 1.0$.
We increment $F_D$ from zero to a maximum value, spending
$2 \times 10^6$ to $5 \times 10^6$ simulation time steps at each
driving force increment in order to obtain a stationary
state average velocity measurement.

\section{Skyrmion Hall Angle at Commensurate and Incommensurate Fillings}

\begin{figure}
\includegraphics[width=3.5in]{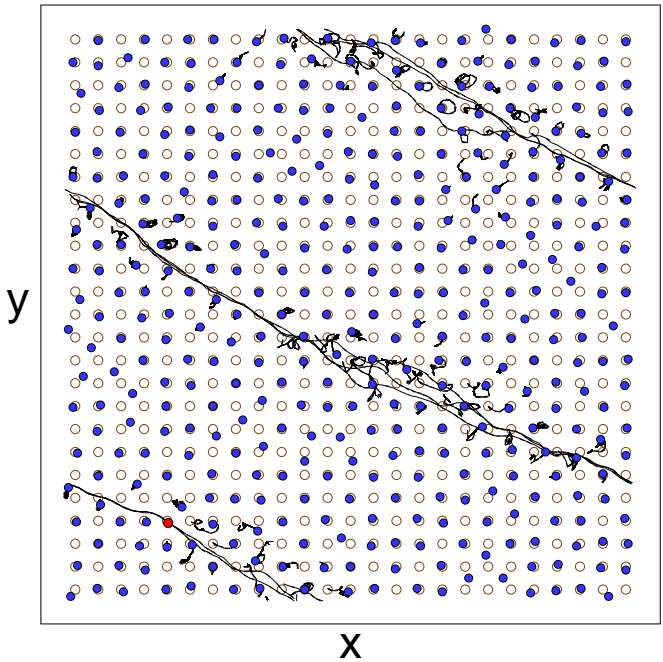}
\caption{ 
Image of a sample containing a square pinning lattice
showing background skyrmions (blue filled circles),
pinning site locations (brown circles), the driven skyrmion (red filled circle), and
the skyrmion trajectories (black lines) during a fixed time window.
Here
$F_{p} = 0.25$,
$\alpha_{m}/\alpha_{d} = 1.0$,
$\theta^{\rm int}_{sk} = -45^\circ$,
and $F_{D} = 1.0$
at an incommensurate filling
of $f = 0.62$ where the
background skyrmions are disordered.
The driven skyrmion moves in a disordered manner along
$\theta_{sk} = -27^\circ$.
	}
\label{fig:1}
\end{figure}

In Fig.~\ref{fig:1} we show
a snapshot of a system containing a square pinning lattice
with $F_{p} = 0.25$,  $\alpha_{m}/\alpha_{d} = 1.0$, $\theta^{\rm int}_{sk} = -45^\circ$,
and $F_{D} = 1.0$
at an incommensurate filling of  $f = 0.62$
where the
background skyrmions are disordered.
The skyrmion follows a disordered trajectory
with
$\theta_{sk} = -27^\circ$,
indicating that the interactions with the pinning and the other skyrmions
have depressed
the magnitude of $\theta_{sk}$ below
that of its intrinsic value $\theta^{\rm int}_{sk}$.

\begin{figure}
\includegraphics[width=3.5in]{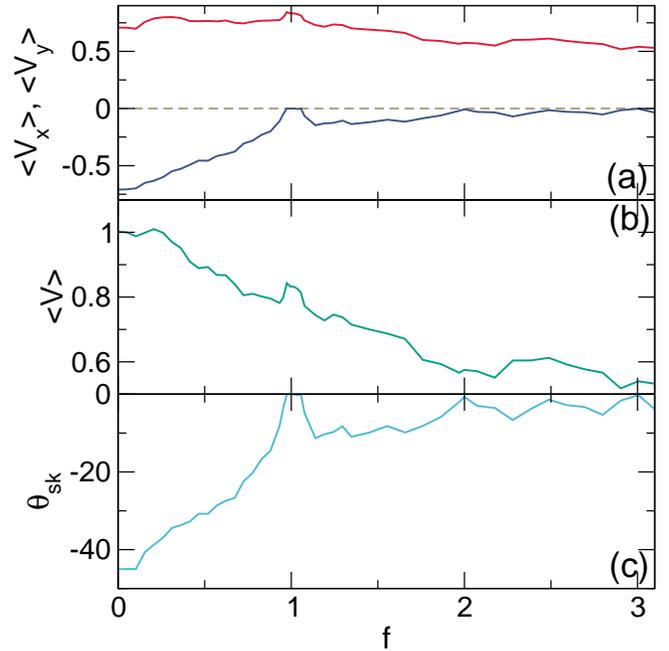}
\caption{Behavior under varied
filling fraction $f=N_s/N_p$ for the system shown in Fig.~\ref{fig:1} with
a square pinning array, $F_p=0.25$, $\alpha_m/\alpha_d=1.0$,
$\theta_{sk}^{\rm int}=-45^\circ$, and $F_D=1.0$.
(a) $\langle V_{x}\rangle$ (red) and $\langle V_{y}\rangle$ (blue) versus
$f$.
(b) $\langle V\rangle = (\langle V_x\rangle^2 + \langle V_y\rangle^2)^{1/2}$
versus $f$. (c) $\theta_{sk} = \arctan(\langle V_{y}\rangle/\langle V_{x}\rangle)$
versus $f$, 
which goes to zero at commensurate fillings $f = 1.0$ and $f=2.0$.  
}
\label{fig:2}
\end{figure}

\begin{figure}
\includegraphics[width=3.5in]{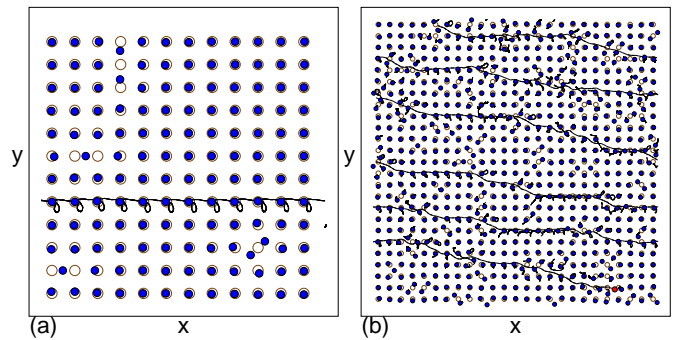}
\caption{
Images of samples containing a square pinning lattice showing background
skyrmions (blue filled circles), pinning site locations (brown circles), the driven
skyrmion (red filled circle), and the skyrmion trajectories during a fixed time window
for the system from Fig.~\ref{fig:1} with $F_p=0.25$, $\alpha_m/\alpha_d=1.0$,
$\theta_{sk}^{\rm int}=-45^\circ$, and $F_D=1.0$.
(a) A portion of the sample at $f = 1.0$, where
$\theta_{sk} = 0^\circ$. Here the driven skyrmion is outside of the visible frame.
(b) The entire sample at $f = 1.14$, where $\theta_{sk} = -10^\circ$.  
	}
\label{fig:3}
\end{figure}

In Fig.~\ref{fig:2}(a) we plot $\langle V_{x}\rangle$ and
$\langle V_{y}\rangle$ versus filling fraction $f$ for the system in Fig.~\ref{fig:1}.
Figure~\ref{fig:2}(b) shows the net skyrmion
velocity $\langle V\rangle$
and Fig.~\ref{fig:2}(c) illustrates the corresponding
$\theta_{sk} = \arctan(\langle V_{y}\rangle/\langle V_{x}\rangle)$.
In the limit $f=0$ where only the driven skyrmion is present,
$\langle V\rangle = 1.0$ and $\theta_{sk} = 45^\circ$.
As $f$ increases, $\langle V_{y}\rangle$ and $\theta_{sk}$
both go to zero at $f = 1.0$, where a peak appears in
both $\langle V_{x}\rangle$ and 
$\langle V\rangle$.
In Fig.~\ref{fig:3}(a) we show a blow up of the skyrmion
motion at $f = 1.0$ for the system in Fig.~\ref{fig:2}.
The background skyrmions
form an ordered commensurate square lattice, while the driven skyrmion
moves along a 1D interstitial channel between two
adjacent pinning rows.
As $f$ is further increased, $\langle V_y\rangle$ and 
$\theta_{sk}$ become finite again over the range $1.0 < f < 2.0$ and
the skyrmion Hall angle becomes finite, as illustrated
in Fig.~\ref{fig:3}(b) 
at $f = 1.14$ where $\theta_{sk} = -10^\circ$.
The magnitude of the skyrmion Hall angle is smaller than that of $\theta_{sk}^{\rm int}$
due to the larger number of skyrmion-skyrmion collisions that occur
at the higher densities.
At $f = 2.0$, $\theta_{sk} = 0.0$ when the
background skyrmions
form an ordered checkerboard state and the
driven skyrmion follows a 1D path along the $x$ direction.
For higher fillings, 
$\langle V\rangle$ and the magnitude of
$\theta_{sk}$ 
gradually decrease.

\begin{figure}
\includegraphics[width=3.5in]{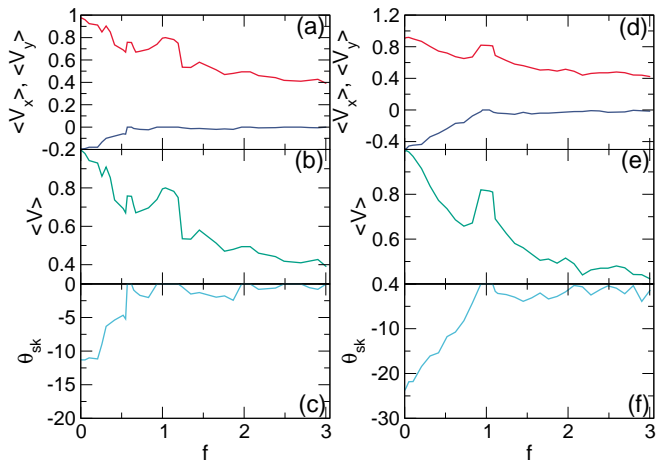}
\caption{
Behavior under varied filling fraction $f$ for the system from Fig.~\ref{fig:2}
with a square pinning array, $F_p=0.25$, and $F_D=1.0$,
but
with
(a,b,c) $\alpha_{m}/\alpha_{d} = 0.204$ and $\theta^{\rm int}_{sk}= -11.54^\circ$
and
(d,e,f) $\alpha_m/\alpha_d=0.436$ and $\theta^{\rm int}_{sk}=-21.9^\circ$.
(a,d) $\langle V_{x}\rangle$ (red) and $\langle V_{y}\rangle$ (blue) versus
$f$.
(b,e) $\langle V\rangle$ versus $f$.
(c,f) $\theta_{sk}$ versus $f$.
	}
\label{fig:4}
\end{figure}

In Fig.~\ref{fig:4}(a,b,c) we plot $\langle V_x\rangle$, $\langle V_y\rangle$,
$\langle V\rangle$, and $\theta_{sk}$
versus $f$ for the system from Fig.~\ref{fig:2} 
but with stronger dissipation of $\alpha_{m}/\alpha_{d} = 0.204$, 
where $\theta^{\rm int}_{sk}= -11.54^\circ$.
We find that $\theta_{sk}$ is equal to
zero at $f=1.0$ and $f=2.0$, and is small in magnitude
for all $f > 0.5$.
There is still a strong peak 
in $\langle V_{x}\rangle$ and $\langle V\rangle$
at $f = 1.0$, with a smaller peak appearing at $f = 2.0$.
In Fig.~\ref{fig:4}(d,e,f), similar behavior appears
at
$\alpha_{m}/\alpha_d = 0.436$ with $\theta^{\rm int}_{sk} = -21.9^\circ$,
where there is a strong peak 
in $\langle V\rangle$ at $f = 1.0$.

\begin{figure}
\includegraphics[width=3.5in]{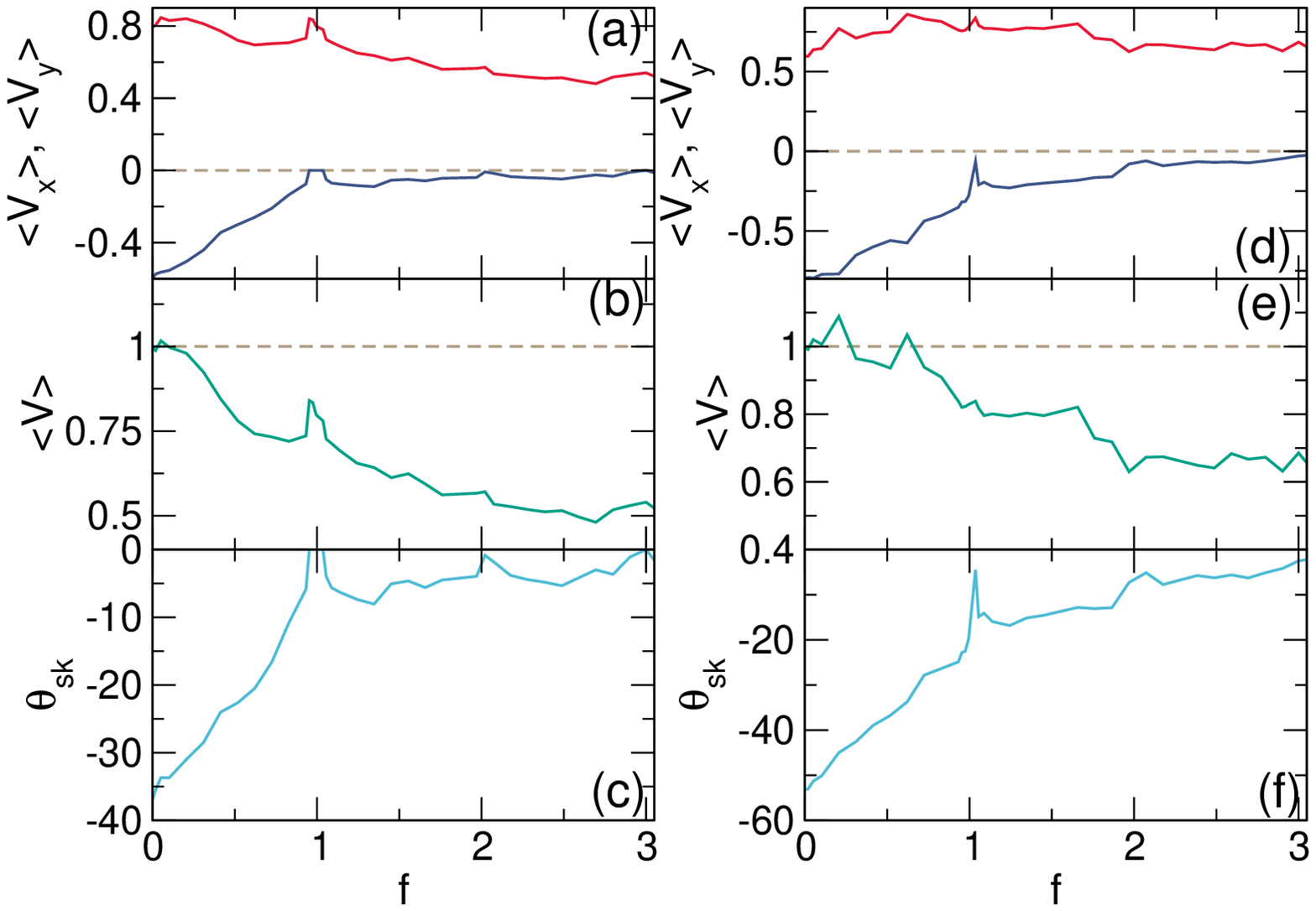}
\caption{
Behavior under varied $f$ for the system from Fig.~\ref{fig:2} with a
square pinning array, $F_p=0.25$, and $F_D=1.0$,
but with
(a,b,c) $\alpha_{m}/\alpha_{d} = 0.57$ and $\theta^{\rm int}_{sk}= -30^\circ$
and (d,e,f) $\alpha_m/\alpha_d=1.33$ and $\theta^{\rm int}_{sk}=-53^\circ$.
(a,d) $\langle V_{x}\rangle$ (red) and $\langle V_{y}\rangle$ (blue) versus
$f$.
(b,e) $\langle V\rangle$ versus $f$.
(c,f) $\theta_{sk}$ versus $f$.
The dashed lines in (b) and (e) are at $\langle V\rangle = 1.0$, 
showing that there is a small boost effect in
panel (e) for $f < 1.0$.  
        }
\label{fig:5}
\end{figure}

The behavior of $\langle V_x\rangle$, $\langle V_y\rangle$, $\langle V\rangle$, and
$\theta_{sk}$ versus $f$ in a system with
$\alpha_{m}/\alpha_{d} = 0.57$ and $\theta^{\rm int}_{sk}= -30^\circ$
appears in Fig.~\ref{fig:5}(a,b,c).
There are peaks in $\langle V\rangle$ at $f = 1.0$, 2.0, and $3.0$,
coinciding with fillings for which
$\theta_{sk} = 0^\circ$.
Figure~\ref{fig:5}(d,e,f) shows the same quantities for a sample with
 $\alpha_{m}/\alpha_{d} = 1.33$
and $\theta^{\rm int}_{sk}= -53^\circ$.
At the matching fillings,
$\langle V_{y}\rangle$ has a dip in magnitude but does not drop completely
to zero. 
There is a cusp in $\theta_{sk}$ to 
$\theta_{sk}=-4^\circ$
at $f = 1.0$, and another more rounded cusp appears
at $f = 2.0$.
For these parameters, there is no peak in $\langle V\rangle$
at $f = 1.0$, but there is
a smaller peak at $f = 0.66$.
In Fig.~\ref{fig:5}(e) we observe
regions over which $\langle V\rangle > 1.0$,
indicated by locations where the curve rises above the dashed line. This is
a signature of
a velocity boost
produced by interactions with the background skyrmions and the pinning
sites.
In general, as the density increases, there are more collisions with the background
skyrmions that increase the effective dissipative velocity; however, since the
collisions can also contribute to the odd viscosity, there is a competition between
the viscosity components. When the odd viscosity term dominates,
the velocity boost effect can emerge
\cite{Reichhardt21a,Reichhardt21c}.

\begin{figure}
\includegraphics[width=3.5in]{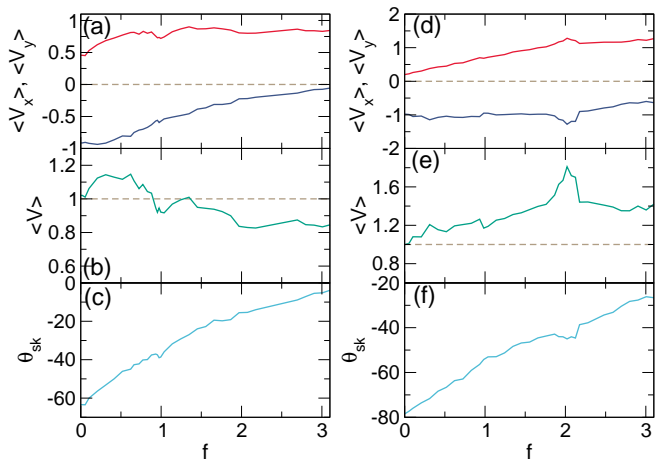}
\caption{
Behavior under varied $f$ for the system from Fig.~\ref{fig:2} with a square
pinning array, $F_p=0.25$, and $F_D=1.0$,
but with  
(a,b,c) $\alpha_{m}/\alpha_{d} = 2.065$ and $\theta^{\rm int}_{sk}= -64^\circ$
and
(d,e,f)  $\alpha_{m}/\alpha_{d} = 4.92$ and $\theta^{\rm int}_{sk}= -78^\circ$.	
(a,d) $\langle V_{x}\rangle$ (red) and $\langle V_{y}\rangle$ (blue) versus $f$.
(b,e) $\langle V\rangle$ versus $f$.
(c,f) $\theta_{sk}$ versus $f$.
The dashed lines in (b) and (e) indicate the value $\langle V\rangle = 1.0$;
boost effects are present when $\langle V\rangle$ rises above this value.
        }
\label{fig:6}
\end{figure}

For a sample with $\alpha_{m}/\alpha_{d} = 2.065$ and $\theta^{\rm int}_{sk}= -64^\circ$,
we plot $\langle V_x\rangle$, $\langle V_y\rangle$, $\langle V\rangle$, and
$\theta_{sk}$ versus $f$ in
Fig.~\ref{fig:6}(a,b,c).
The peaks that appeared in $\langle V\rangle$
at $f = 1.0$ and $f=2.0$ for lower $\alpha_m/\alpha_d$ have not only disappeared,
but there is now
a dip in $\langle V\rangle$ at $f = 1.0$ accompanied by
a small increase in the magnitude of $\theta_{sk}$.
A velocity boost appears over the entire range
$0.1 < f < 1.0$, as indicated by the curve rising above the dashed $\langle V\rangle=1.0$
line
in Fig.~\ref{fig:6}(b).
There is an increase in $\langle V_x\rangle$ over the same range of fillings
since the boosted velocity is aligned with the driving or $+x$ direction,
and, when the Magnus force is large enough,
the magnitude of the boost increases as the magnitude of the skyrmion Hall
angle $\theta_{sk}$ decreases \cite{Reichhardt21a}.
We plot the same quantities for a sample with
$\alpha_{m}/\alpha_{d} = 4.92$ and $\theta^{\rm int}_{sk}= -78^\circ$ in
Fig.~\ref{fig:6}(d,e,f). 
Here there is no large feature
at $f = 1.0$; instead,
strong peaks or dips appear in
$\langle V_{x}\rangle$, $\langle V_{y}\rangle$ and $\langle V\rangle$
at $f = 2.0$.  For this filling,
$\theta_{sk} = -45^\circ$,
indicating that the commensurate structure is directionally locked with one of
the major symmetry angles of the pinning array.
As $f$ increases further, the magnitude of $\theta_{sk}$
resumes its decrease after passing through the locking step at $\theta_{sk}=-45^\circ$.
At $f = 2.0$, the system forms an ordered checkerboard 
state with a well defined symmetry direction; however, away from $f = 2.0$, the
structure becomes disordered again and the directional locking is lost. 
In Fig.~\ref{fig:6}(d), $\langle V_{x}\rangle$ increases with increasing
$f$ while
the magnitude of $\langle V_{y}\rangle$ undergoes a moderate decrease.
Figure~\ref{fig:6}(e) shows that a velocity boost is present over the entire range of $f$
except at very small values of $f$, with the largest boost appearing at
$f = 2.0$. At this filling, $\langle V\rangle \approx 1.8$,
nearly twice as large as the expected velocity in the limit of a single skyrmion
and no substrate.

Taken together, the results
in this section indicate that commensuration effects can 
both reduce the skyrmion Hall angle and
also speed up the skyrmion motion, two features that are desirable for
applications.
For values of $\alpha_{m}/\alpha_{d}$ higher than what is shown here,
we observe similar trends;
however, as the relative strength of the damping term becomes small,
the motion becomes increasingly
disordered and fluid-like,
the curves become smooth, and the directional locking effects disappear.
In addition, for values of $f$ higher than those shown here,
the velocity boost is eventually destroyed.
For lower $F_{D}$ and high fillings, a jamming effect can occur in which
the driven skyrmion
becomes pinned or jammed due to its strong interactions with the background
skyrmions,
similar to
what is found in other active rheology systems \cite{Habdas04,Reichhardt10}. 

\subsection{Fluctuations and Noise}

\begin{figure}
\includegraphics[width=3.5in]{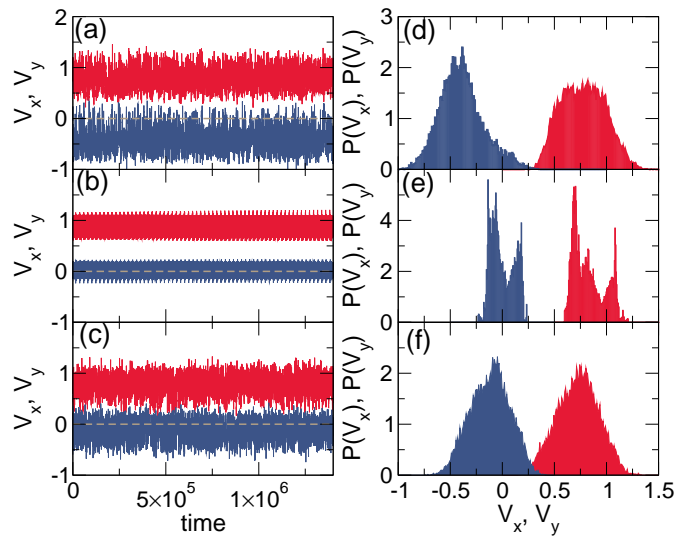}
\caption{
Velocity fluctuation data for the system from Fig.~\ref{fig:2}  with a square
pinning array, $F_p=0.25$, $F_D=1.0$, $\alpha_m/\alpha_d=1.0$,
and $\theta_{sk}^{\rm int}=-45^\circ$ at (a,d) $f=0.62$, (b,e) $f=1.0$, and (c,f) $f=1.14$.
(a,b,c) Time series of the instantaneous velocities $V_x$ (red) and $V_y$ (blue).
(d,e,f) The corresponding velocity distributions $P(V_x)$ (red) and $P(V_y)$ (blue).
The distributions have a Gaussian character at $f=0.62$ in (d) and $f=1.14$ in (f)
when the motion is disordered, but develop sharp peaks at $f=1.0$ in (e) when the
motion is periodic.
	}
\label{fig:7}
\end{figure}

We can also characterize the behavior
of the commensurate and non-commensurate states
using the fluctuation properties of 
the time series of the driven skyrmion velocity.
In general, we find ordered motion at commensurate fillings for which
$\theta_{sk} = 0^\circ$ and disordered motion at
incommensurate fillings where $\theta_{sk}$ is finite. 
In Fig.~\ref{fig:7}(a,b,c) we plot the time series of the instantaneous velocities
$V_x$ and $V_y$
for the system in Fig.~\ref{fig:2} 
with $\theta^{\rm int}_{sk} = -45^\circ$.
At $f=0.62$ in Fig.~\ref{fig:7}(a),
where $\theta_{sk} = -28^\circ$,  the velocities fluctuate rapidly and, as
shown in the
corresponding velocity distribution plots $P(V_x)$ and $P(V_y)$ in Fig.~\ref{fig:7}(d),
the distributions have a broadened Gaussian form that is characteristic of
random fluctuations.
At $f=1.0$ in Fig.~\ref{fig:7}(b,e),
the motion is periodic 
and $P(V_{y})$ is centered at zero.
Here the
velocity distributions are non-Gaussian and have sharp features indicative of the
repeated periodic motion.
In Fig.~\ref{fig:7}(c,f) at $f = 1.14$, 
corresponding to the motion
with $\theta_{sk} = -10^\circ$
illustrated in Fig.~\ref{fig:3}(b),
the flow is once again random with
Gaussian velocity
distributions.
We find similar
behaviors for commensurate and incommensurate fillings at
other values of $\alpha_{m}/\alpha_{d}$,
including the $\langle V_y\rangle=0$ states
at $f = 1.0$ and $f=2.0$ where the motion is periodic. 
For higher values of $\alpha_{m}/\alpha_{d}$
such as
$\alpha_{m}/\alpha_{d} = 1.33$ in Fig.~\ref{fig:5}(f), 
where $\theta_{sk}$ does not reach zero
but is strongly reduced
at $f = 1.0$ and $f=2.0$,
the motion is mostly periodic with
occasional jumps in the transverse direction. 

\begin{figure}
\includegraphics[width=3.5in]{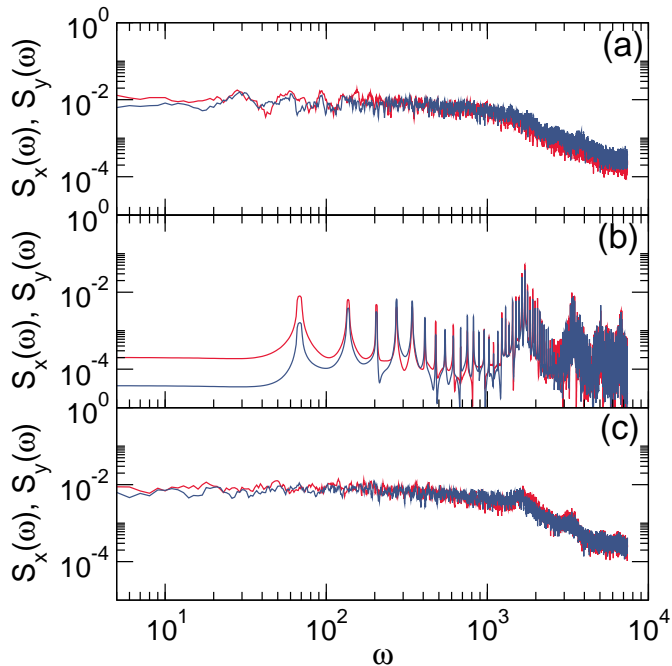}
\caption{
The power spectra $S_x(\omega)$ (red) and $S_y(\omega)$ (blue)
for the velocity time series $V_x$ and $V_y$, respectively,
in Fig.~\ref{fig:7}
from samples with a square pinning array, $F_p=0.25$, $F_D=1.0$, 
$\alpha_m/\alpha_d=1.0$, and $\theta_{sk}^{\rm int}=-45^\circ$. 
(a) $f = 0.62$.
(b) $f=1.0$, where there is a strong narrow band noise signature.
(c) $f=1.14$.
        }
\label{fig:8}
\end{figure}

The skyrmion motion
can also be characterized by examining the velocity noise,
as considered in both simulations \cite{Diaz17} and experiments \cite{Sato19}.
From the time series in Fig.~\ref{fig:7}, we 
can extract the power spectrum
$S_{\alpha}(\omega) = |\int V_{\alpha}(t)e^{-i\omega t} dt|^2$, where $\alpha=x, y$.  
In Fig.~\ref{fig:8} we plot the power spectra from the time series of both
$V_x$ and $V_y$ for the system in Fig.~\ref{fig:7}.
At $f = 0.62$ in Fig.~\ref{fig:8}(a) there are
no peaks in the power spectra,
indicating that the motion is random,
while at $f = 1.0$ in Fig.~\ref{fig:8}(b), there is a strong narrow
band noise signal as indicated by the sequence of peaks.
In this case there are two overlapping periodic signals.
The first, at higher frequencies, arises from the periodic  motions of the
driven skyrmion as it interacts
with the ordered commensurate pinned skyrmions. The
second, at lower frequencies, is produced by
a small number of defects that are present in the commensurate configuration, which
generate a time-of-flight velocity signature.
For the disordered motion at $f = 1.14$ in Fig.~\ref{fig:8}(c),
the noise signature has a broad band character.
There is a weak periodic  signal at higher frequencies due to the low
value of $\theta_{sk}$ at this filling, which causes the skyrmion to flow through
an interstitial channel between two adjacent pinning rows with
infrequent but roughly periodic hops occurring from one interstitial channel
to the next.

\begin{figure}
\includegraphics[width=3.5in]{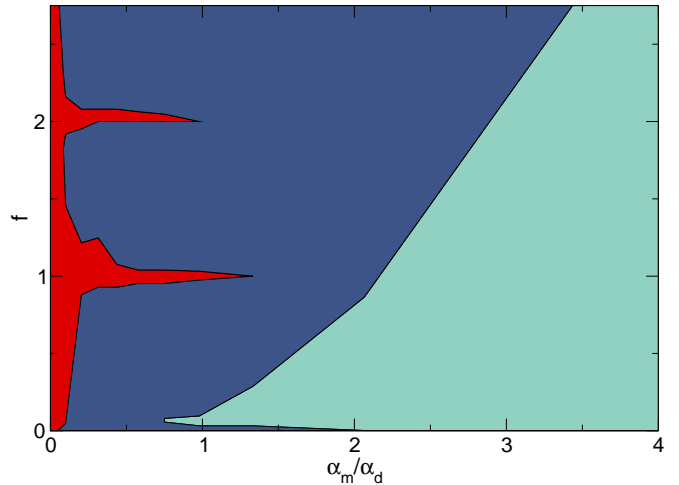}
\caption{
Dynamic phase diagram as a function of filling fraction $f$ versus
$\alpha_m/\alpha_d$
for the system in Fig.~\ref{fig:2} with a square pinning array, $F_p=0.25$, and $F_D=1.0$
constructed using the features in the transport curves and
the behavior of $\theta_{sk}$.
In the red region, $\theta_{sk}=0.0^\circ$, a condition that extends out to
higher $\alpha_m/\alpha_d$ at commensurate fillings.
In the blue region, $\theta_{sk}$ is finite but there is no
velocity boost, while in the green region,
$\theta_{sk}$ remains finite and a velocity boost appears.
	}
\label{fig:9}
\end{figure}

In Fig.~\ref{fig:9} we
use the
behavior of the transport curves and $\theta_{sk}$ to construct a dynamic
phase diagram as a function of filling fraction $f$ versus $\alpha_m/\alpha_d$
for the system in Fig.~\ref{fig:2}.
In the red region, the skyrmion moves strictly along the direction of drive
and $\theta_{sk} = 0^\circ$.
In the blue region, the skyrmion Hall angle is finite but there is no velocity boost,
and in the green region, there is both a finite skyrmion Hall angle and a finite
velocity boost.
Near the commensurate conditions
of $f = 1.0$ and $f=2.0$, the window of zero skyrmion Hall angle extends out
to larger values of $\alpha_m/\alpha_d$.
For these same commensurate conditions, even when the value of $\theta_{sk}$
becomes finite, it is still suppressed relative to its value away from
commensuration, a feature that is not illustrated in the plot.

\section{Velocity Force Curves}

\begin{figure}
\includegraphics[width=3.5in]{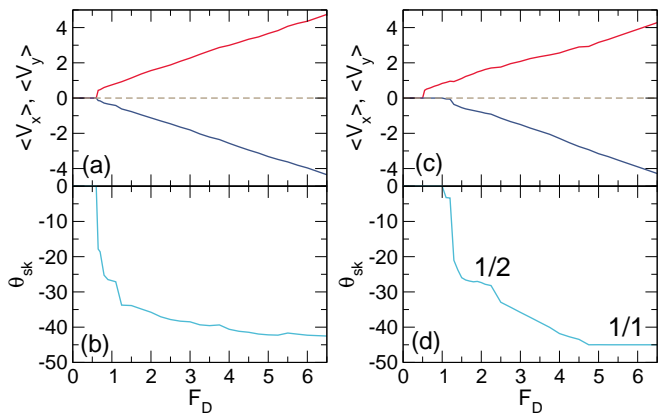}
\caption{
(a)  $\langle V_{x}\rangle$ (red) and $\langle V_{y}\rangle$ (blue)
versus driving force $F_{D}$ for the system
from Fig.~\ref{fig:2} with a square pinning array, $F_p=0.25$,
$\alpha_m/\alpha_d=1.0$, and $\theta^{\rm int}_{sk}=-45^\circ$
at a filling fraction of $f = 0.62$. 
(b) The corresponding $\theta_{sk}$ versus $F_{D}$.
(c) $\langle V_x\rangle$ (red) and $\langle V_y\rangle$ (blue) versus $F_D$
for the same system at $f=1.0$,
where there is a two step depinning process.
(d) The corresponding $\theta_{sk}$ versus $F_D$ contains
two directional locking steps along the angles
$\theta_{sk}=\arctan(1/2)$ and $\theta_{sk}=\arctan(1/1)$.
	}
\label{fig:10}
\end{figure}

We next consider the effect of varying the driving force
$F_D$ on the driven particle in the system from Fig.~\ref{fig:2}.
In Fig.~\ref{fig:10}(a) we plot $\langle V_{x}\rangle$ and
$\langle V_{y}\rangle$ versus $F_{D}$
at $f = 0.62$.
There is a single depinning threshold at $F_D = 0.6$,
above which motion occurs simultaneously in both
directions.
Figure~\ref{fig:10}(b) shows the corresponding $\theta_{sk}$
versus $F_D$, where
$\theta_{sk}=0^\circ$ at the depinning threshold.
As $F_D$ increases, the magnitude of $\theta_{sk}$ gradually increases until
it approaches the intrinsic value $\theta_{sk}^{\rm int}=-45^\circ$ at higher
drives.
In Fig.~\ref{fig:10}(c), $\langle V_x\rangle$ and $\langle V_y\rangle$ versus
$F_D$ for a sample with $f=1.0$ reveal the presence of
a two step  depinning threshold.
The first depinning transition occurs at
$F_{D} = 0.55$, and for
$0.55 < F_{D} < 1.25$
the motion is only along the $x$-direction with $\theta_{sk}=0^\circ$.
The second depinning transition occurs
at $F_{D} = 1.25$, above which
finite motion in the $y$ direction appears.
In the corresponding plot of $\theta_{sk}$ versus $F_D$ in
Fig.~\ref{fig:10}(d),
$\theta_{sk}$ is zero above the first depinning transition
and begins to increase in magnitude
above the second depinning transition. 
There are also several 
steps or plateaus in the velocity, including
one with $\theta_{sk} \sim -27^\circ$ and another with $\theta_{sk}=-45^\circ$. These
correspond to directional locking steps
with $\theta_{sk}=\phi=\arctan(n/m)$.
The $n/m=1/2$ step has $\phi=\arctan(1/2)=-25.56^\circ$,
while the $1/1$ step has $\phi=-45^\circ$.

\begin{figure}
\includegraphics[width=3.5in]{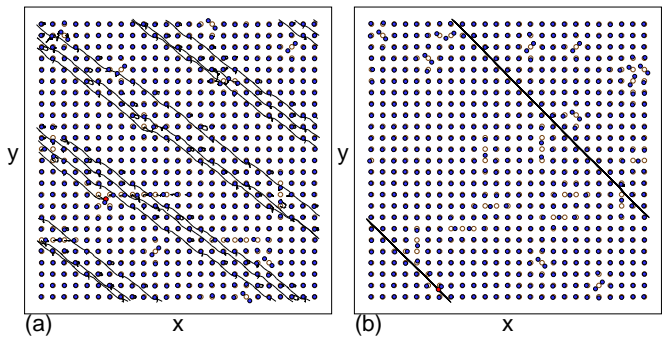}
\caption{
Images of samples containing a square pinning lattice showing background
skyrmions (blue filled circles), pinning site locations (brown circles), the driven
skyrmion (red filled circle), and the skyrmion trajectories during a fixed time
window
for the system in Fig.~\ref{fig:10}(c,d) with $F_p=0.25$, $\alpha_m/\alpha_d=1.0$,
and $\theta^{\rm int}_{sk}=-45^\circ$
at a filling fraction of
$f = 1.0$.
(a)  $F_{D} = 2.75$ near the $1/2$ step. (b)  $F_{D} = 6.0$ along the $1/1$ step.
	}
\label{fig:11}
\end{figure}

In Fig.~\ref{fig:11}(a) we illustrate the skyrmion trajectories for the sample in
Fig.~\ref{fig:10}(c,d) at a non-step drive of
$F_{D} = 2.75$, which is close to but not on the $1/2$ step.
There is a combination of ordered and disordered motions,
with regions in which
the skyrmion translates by a distance
$2a$ in the $x$ direction and $-a$ in
the $y$ direction,
combined with regions in which the skyrmion makes
occasional jumps in other
directions. 
Figure~\ref{fig:11}(b) shows the same system at $F_D=6.0$ on the
$1/1$ step,
where
the skyrmion moves in an orderly periodic fashion strictly along $-45^\circ$.

\begin{figure}
\includegraphics[width=3.5in]{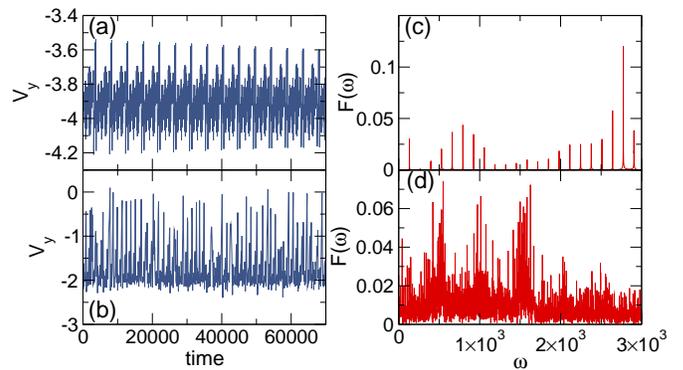}
\caption{
(a,b) The time series of $V_y$ and (c,d) its corresponding Fourier transform $F(\omega)$
for the system from Fig.~\ref{fig:10}(c,d) with a square pinning array, $F_p=0.25$,
$\alpha_m/\alpha_d=1.0$, and $\theta_{sk}^{\rm int}=-45^\circ$
at
a filling fraction of $f=1.0$.
(a,b) $F_{D} = 6.0$ on the $1/1$ locking step where the motion is
periodic.
(c,d) A non-step region at $F_{D} = 3.25$
where the flow is more disordered and the peaks in $F(\omega)$ are broadened.
	}
\label{fig:12}
\end{figure}

In Fig.~\ref{fig:12}(a) we plot the time series of $V_{y}$
on the $1/1$ step
at $F_{D} = 6.0$
in the system from Fig.~\ref{fig:10}(c,d),
while in Fig.~\ref{fig:12}(c) we show the corresponding
Fourier transform $F(\omega)=\int V_y e^{-i\omega t}dt$, where there is
a strong narrow band signal.
The time series of $V_y$ for a non-step region at $F_{D} = 3.25$
with partially periodic motion
appears
in Fig.~\ref{fig:12}(b), and the broadened signature of
the corresponding $F(\omega)$
is shown in Fig.~\ref{fig:12}(d).
In general, 
as a function of increasing $F_{D}$
we find that in
regimes of directional locking where there are steps
in $\theta_{sk}$, the periodic motion produces narrow band velocity noise,
while in regimes where no steps are present,
the motion is more disordered and the velocity noise is either
broad band
or contains weakly periodic signals. 

For individual skyrmions moving over a square substrate
in the absence of background skyrmions, previous work 
showed that the velocity-force curve and skyrmion
Hall angle exhibit
a series of steps corresponding to directional locking
centered on the values
$\theta_{sk} = \arctan(n/m)$ with $m,n$ integer
\cite{Reichhardt15a}.
For the active rheology situation we consider here,
due to the large number of collisions
the driven skyrmion experiences with background skyrmions,
only the most prominent locking steps are visible, as
illustrated in Fig.~\ref{fig:10}(d).
In general, higher order
directional locking only occurs at the commensurate conditions
of $f = 1.0$ and $f = 2.0$,
but there is some weaker higher order locking at
some incommensurate fillings as well. The number of
accessible steps also depends on the value of $\alpha_{m}/\alpha_{d}$;
as this quantity becomes larger,
a greater number of steps beyond $1/1$ can 
be accessed. 

\begin{figure}
\includegraphics[width=3.5in]{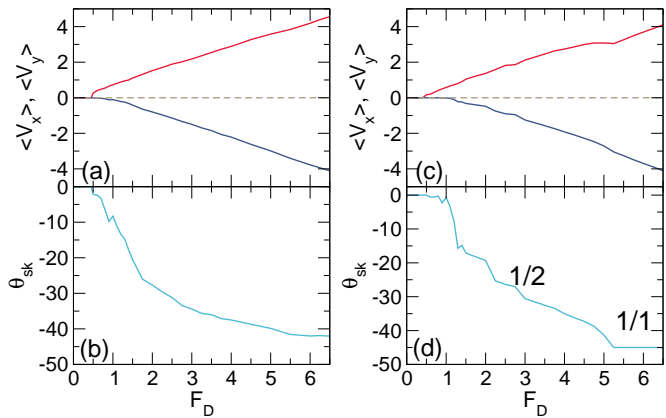}
\caption{
(a)   $\langle V_{x}\rangle$ (red) and $\langle V_{y}\rangle$ (blue) versus
$F_{D}$ for the system from Fig.~\ref{fig:2} with a square pinning array,
  $F_p=0.25$, $\alpha_m/\alpha_d=1.0$, and $\theta^{\rm int}_{sk}=-45^{\circ}$
at $f = 1.1392$.
(b) The corresponding $\theta_{sk}$ versus $F_{D}$.
(c) $\langle V_x\rangle$ (red) and $\langle V_y\rangle$ (blue) versus $F_D$ for
the same system at
$f = 2.0$ where there is a two step depinning process.
(d) The corresponding $\theta_{sk}$ versus $F_D$ contains
two directional locking steps at $1/1$ and $1/2$.
        }
\label{fig:13}
\end{figure}

\begin{figure}
\includegraphics[width=3.5in]{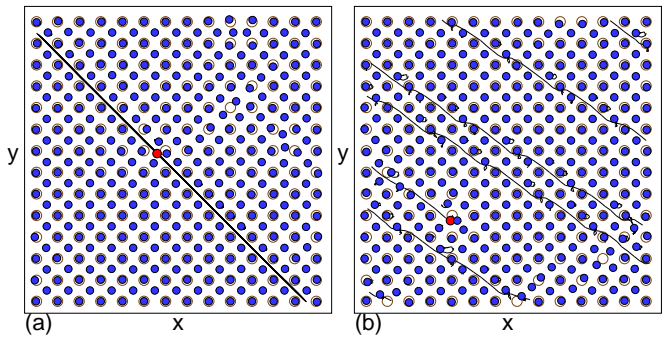}
\caption{
Images of portions of samples containing a square pinning lattice showing background
skyrmions (blue filled circles), pinning site locations (brown circles), the driven
skyrmion (red filled circle), and the skyrmion trajectories during a fixed time window
for the system in Fig.~\ref{fig:13}(c,d) with $F_p=0.25$, $\alpha_m/\alpha_d=1.0$,
and $\theta_{sk}^{\rm int}=-45^\circ$
at
$f=2.0$.
(a)  $F_{D} = 6.0$ on the $1/1$ step.
(b)  $F_{D} = 3.25$ in a non-step region.  
        }
\label{fig:14}
\end{figure}

In Fig.~\ref{fig:13}(a,b) we plot $\langle V_x\rangle$, $\langle V_y\rangle$, and
$\theta_{sk}$ versus $F_D$ for the system from
Fig.~\ref{fig:10}
at a filling fraction of $f = 1.1392$.
The locking step features are reduced or absent,
while  $\langle V_{x}\rangle$ and
$\langle V_{y}\rangle$ both
become finite at a single depinning transition near $F_{D} = 0.5$.
Figure~\ref{fig:13}(c,d) shows the
same quantities for 
$f = 2.0$, where there is a two step depinning 
transition in which $\langle V_{x}\rangle$ becomes finite at $F_{D} = 0.5$ 
and $\langle V_{y}\rangle$ does not become finite until $F_{D} = 1.15$.
There is a reduced $1/2$ locking step, but the $1/1$ locking step
remains robust.
In Fig.~\ref{fig:14}(a) we illustrate the skyrmion trajectories in a subsection
of the sample
at $F_D=6.0$ on the $1/1$ locking step,
where the driven skyrmion moves at $-45^\circ$ through an interstitial channel
in the checkerboard lattice formed by the background skyrmions.
The $1/1$ step is associated with a cusp in $\langle V_x\rangle$ and
$\langle V_y\rangle$ as indicated in Fig.~\ref{fig:13}(c).
For driving individual skyrmions over a periodic substrate
in the absence of background skyrmions,
similar cusps in the velocity-force curves appear
at several of the transitions into and out of the directional
locking steps \cite{Reichhardt15a}. 
Figure~\ref{fig:14}(b) shows the same
system in a non-step region at $F_{D} = 3.25$,
where the motion is more disordered and $\theta_{sk}$ is smaller
in magnitude.

\section{Varied Pinning Strength}

\begin{figure}
\includegraphics[width=3.5in]{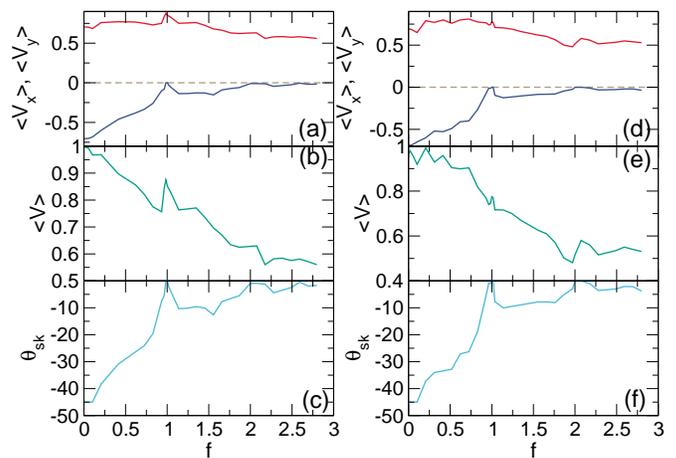}
\caption{Behavior under varied filling fraction $f$ for the system shown in
Fig.~\ref{fig:2} with a square pinning array, $\alpha_m/\alpha_d=1.0$,
$\theta_{sk}^{\rm int}=-45^\circ$, and $F_D=1.0$.
(a,d) $\langle V_x\rangle$ (red) and $\langle V_y\rangle$ (blue) versus $f$.
(b,e) $\langle V\rangle$ versus $f$.
(c,f) $\theta_{sk}$ versus $f$.
The pinning strength is (a,b,c) $F_p=0.125$ and (d,e,f) $F_p=0.5$.
}
\label{fig:15}
\end{figure}

We next consider the effect of increasing the pinning strength for the
system in Fig.~\ref{fig:2} with fixed $\alpha_{m}/\alpha_{d} = 1.0$ and $F_{D} = 1.0$.  
In Fig.~\ref{fig:15}(a,b,c) we plot $\langle V_x\rangle$,
$\langle V_y\rangle$, $\langle V\rangle$, and $\theta_{sk}$  
versus $f$ for
a sample with $F_{p}= 0.125$, which is half the value of $F_p$ used in Fig.~\ref{fig:2}.
The general trends remain unchanged, with the magnitude of
$\theta_{sk}$ dropping to zero
at $f = 1.0$ and reaching a value close to zero
at $f = 2.0$.
Figure~\ref{fig:15}(d,e,f) shows the same
quantities in a sample with stronger pinning, $F_{p} = 0.5$.
The shapes of the curves remain similar to those at lower $F_p$, but
there is now a finite window of $f$ near $f=2.0$ where 
$\theta_{sk}=0^\circ$.

\begin{figure}
\includegraphics[width=3.5in]{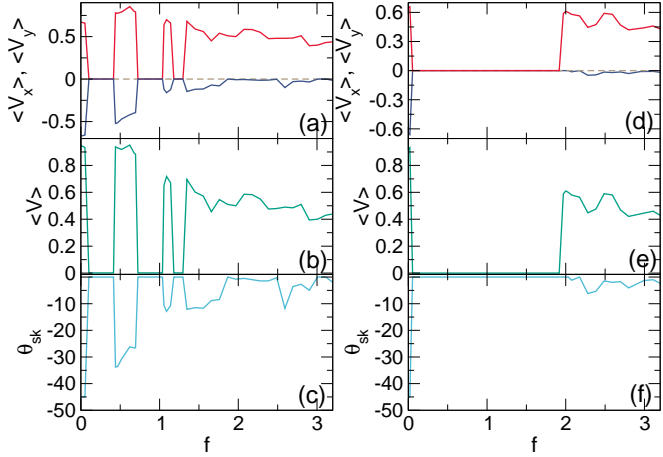}
\caption{Behavior under varied filling fraction $f$ for the system
from Fig.~\ref{fig:2} with a square pinning array, $\alpha_m/\alpha_d=1.0$,
$\theta_{sk}^{\rm int}=-45^\circ$, and $F_D=1.0$.
(a,d) $\langle V_x\rangle$ (red) and $\langle V_y\rangle$ (blue) versus $f$.
(b,e) $\langle V\rangle$ versus $f$.
(c,f) $\theta_{sk}$ versus $f$.
The pinning strength is (a,b,c) $F_p=0.625$ and (d,e,f) $F_p=0.75$.
}
\label{fig:16}
\end{figure}

In Fig.~\ref{fig:16}(a,b,c) we plot $\langle V_{x}\rangle$, $\langle V_{y}\rangle$,
$\langle V\rangle$, and $\theta_{sk}$ versus $f$ for the
same system in Fig.~\ref{fig:15} but at stronger pinning of $F_{p}= 0.625$. 
There are now regions, such as near $f=1.0$, where the system is
pinned.
There are also extended regions near $f = 2.0$
and $f = 3.0$ where
$\theta_{sk}$ is close to zero.
The pinned phase arises
from the interaction between the driven skyrmion and the pinning sites,
which occurs
both directly when the driven skyrmion encounters
a pinning site, and indirectly when the driven skyrmion
experiences a repulsion from a pinned skyrmion.
At low $f$, the driven particle has only direct interactions with the
pinning sites,
and as long as $F_{D}/F_{p} > 1.0$,
it will not become pinned. 
At large $f$, all of the pinning sites are occupied by background skyrmions
and the driven skyrmion never encounters a pinning site directly;
however, for certain incommensurate fillings at which the positions of the
background skyrmions become disordered,
the driven skyrmion
can become pinned
via interactions with skyrmions located at pinning sites.  
The plots of $\langle V_x\rangle$, $\langle V_y\rangle$,
$\langle V\rangle$, and $\theta_{sk}$ versus $f$ in
Fig.~\ref{fig:16}(d,e,f) for the same system at $F_{p} = 0.75$ show
that motion only occurs for $f <  0.05$ and $f > 2.0$ with
a broad pinned window appearing in between those fillings,
while the magnitude of $\theta_{sk}$ remains below $10^\circ$ for the
higher fillings. 

\begin{figure}
\includegraphics[width=3.5in]{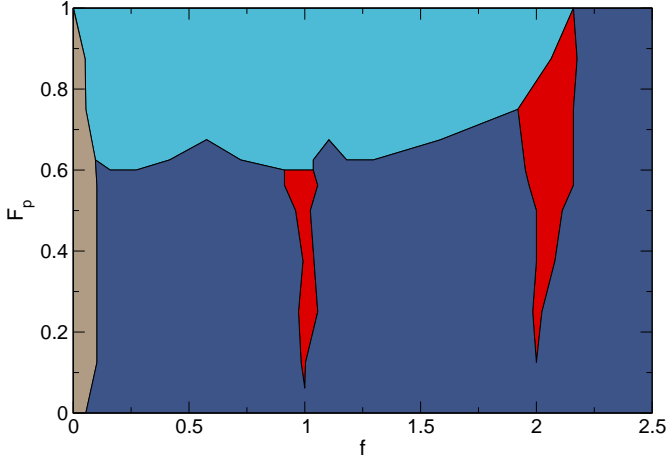}
\caption{Dynamic phase diagram as a function of pinning strength $F_p$ and
filling fraction $f$ for the system in Figs.~\ref{fig:15} and \ref{fig:16} with a
square pinning array, $F_D=1.0$, and $\alpha_m/\alpha_d=1.0$.
The system is pinned in the light blue region,
moving with $\theta_{sk}=0.0^\circ$ in the red region,
moving with $\theta_{sk}=-45^\circ$ in the brown region,
and moving with finite $\theta_{sk}$ and no velocity boost in the
dark blue region.
        }
\label{fig:17}
\end{figure}

Using the features in the transport curves and
$\theta_{sk}$, in Fig.~\ref{fig:17} we construct a dynamic phase diagram as a function
of $F_p$ versus $f$
for the system in Figs.~\ref{fig:15} and \ref{fig:16}
with $\alpha_{m}/\alpha_{d} = 1.0$ and $F_{D} = 1.0$.
A large pinned region appears at larger $F_p$.
There are some smaller pinned regimes (not shown)
above $f = 2.1$.
The intervals of pinning also depend strongly on $F_{D}$.
When $\theta_{sk} \approx 0^\circ$, the motion is locked to the $x$
direction, shown as red regions.
For low $f$ where there are few interactions with background
skyrmions, the brown
region indicates that
the motion is locked to $\theta_{sk}=45^\circ$.
At these low values of $f$, pinned states occur only when
$F_{p} > F_D = 1.0$.
In the blue region, $\theta_{sk}$ is finite and there is no boost effect.
Near the border between the flowing and pinned regimes,
there are some windows of stick-slip motion
which produce
$1/f$ velocity fluctuation noise
and a bimodal velocity distribution with a peak
at $\langle V\rangle = 0.0$ and a second peak at higher velocities.

\begin{figure}
\includegraphics[width=3.5in]{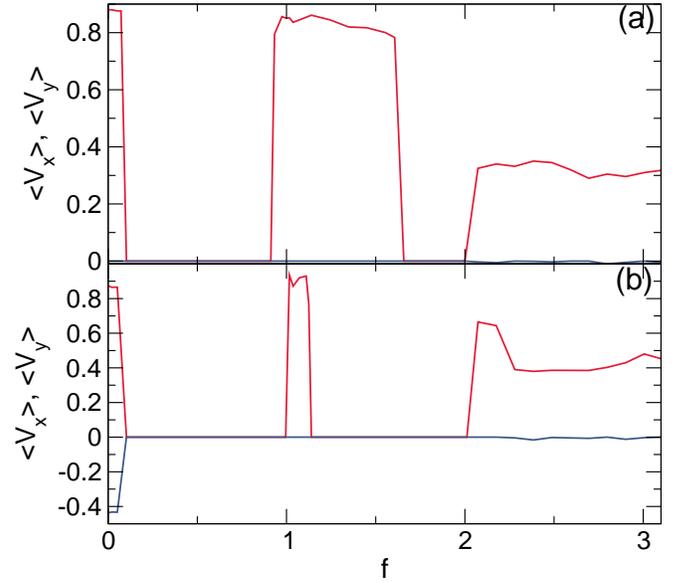}
\caption{
$\langle V_x\rangle$ (red) and $\langle V_y\rangle$ (blue) versus $f$  
  for a system with a square pinning array, $F_p=0.75$, and $F_D=1.0$.
(a) $\alpha_{m}/\alpha_{d} = 0.1$ and $\theta_{sk}^{\rm int}=-5.74^\circ$. 
(b) $\alpha_{m}/\alpha_{d} = 0.5$ and $\theta_{sk}^{\rm int}=-30^\circ$. 
}
\label{fig:18}
\end{figure}

We next consider a sample with strong pinning of $F_p=0.75$ and $F_D=1.0$
where we vary $\alpha_{m}/\alpha_{d}$.
In Fig.~\ref{fig:18}(a)
we plot  $\langle V_{x}\rangle$ and $\langle V_{y}\rangle$ versus $f$
for a system with $\alpha_{m}/\alpha_{d}= 0.1$ and $\theta^{\rm int}_{sk} = -5.74^\circ$.
In this case, $\langle V_{y}\rangle=0$
over the entire range of $f$ measured.
There are two pinned intervals
with $\langle V_x\rangle=0$ at $0.075 < f < 0.92$ 
and $1.6 < f < 2.07$.
For low $f$, the driven skyrmion moves along the pinning rows 
in the $x$-direction. As $f$ increases, the driven skyrmion
begins to collide with other skyrmions and  becomes
pinned by the combination of the pinning and the interactions with the background
skyrmions.
Near $f = 1.0$ 
where the system is more ordered,
the driven skyrmion channels
along the $x$ direction between the pinned skyrmions.
Near $f  = 0.8$, the background skyrmions are disordered enough that
the driven skyrmion trajectory begins to divert into the $y$ direction, but
the driven skyrmion quickly becomes trapped among the background skyrmions.
For $f > 2.0$, there
are regions where the driven skyrmion depins and shepherds some of the
background skyrmions along the $x$ direction.
Figure~\ref{fig:18}(b) shows $\langle V_x\rangle$ and $\langle V_y\rangle$ 
for a sample with
$\alpha_{m}/\alpha_{d} = 0.58$ and $\theta^{\rm int}_{sk} = -30^\circ$.
For low $f$, the skyrmion Hall angle is finite.
There is also  a reduced window
near $f = 1.0$ where the driven skyrmion channels along the $x$ direction.
Figure~\ref{fig:18} demonstrates that
increasing the Magnus component of the dynamics
enhances the
effectiveness of the pinning, a behavior opposite from what is typically observed in
systems with random pinning
\cite{Reichhardt21}.
The high Magnus force causes the driven skyrmion to attempt to move
partially in the $y$ direction rather than strictly along the $x$ direction,
and since the direction of motion no longer coincides with a symmetry direction
of the pinning lattice, encounters with pinning and pinned skyrmions happen
more frequently and increase the effectiveness of the pinning.

\begin{figure}
\includegraphics[width=3.5in]{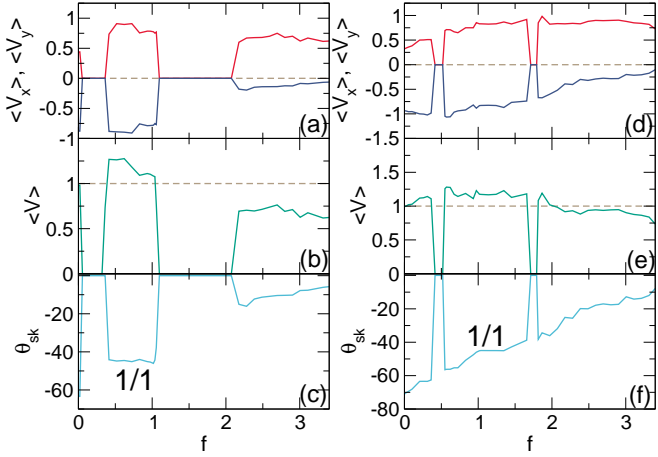}
\caption{
Behavior under varied filling fraction $f$ for a system with a square pinning array,
$F_p=0.75$, and $F_D=1.0$.
(a,d) $\langle V_x\rangle$ (red) and $\langle V_y\rangle$ (blue) versus $f$.
(b,e) $\langle V\rangle$ versus $f$.
(c,f) $\theta_{sk}$ versus $f$.
(a,b,c) $\alpha_m/\alpha_d=2.06$ and $\theta_{sk}^{\rm int}=-64.15^\circ$.
(d,e,f) $\alpha_{m}/\alpha_d = 3.04$ and $\theta_{sk}^{\rm int} = -71.8^\circ$.
        }
\label{fig:19}
\end{figure}

In Fig.~\ref{fig:19}(a,b,c) we plot $\langle V_x\rangle$, $\langle V_y\rangle$,
$\langle V\rangle$, and $\theta_{sk}$ versus $f$ for a sample with
 $\alpha_{m}/\alpha_{d} = 2.06$ and $\theta^{\rm int}_{sk} = -64.15^\circ$. 
From $0.41 < f < 1.08$, there is an unpinned region with motion 
locked to $-45^\circ$, as indicated by the $1/1$ label in
Fig.~\ref{fig:19}(c). 
There is a pinned region for $1.08 < f < 2.16$
followed by another region in which motion occurs in both the $x$ and $y$
directions.
Notice that when $f = 0$,
$\langle V_{x}\rangle = 0.44$,
but that in the range
$0.41 < f < 1.0$,
$\langle V_{x}\rangle \approx 0.9$,
indicating a substantial boost of the velocity in the direction of driving.
There is also a boost in the overall
velocity for $1.08 < f < 2.16$,
as shown in Fig.~\ref{fig:19}(b) where $\langle V\rangle$ rises above the
dashed line marking $\langle V\rangle=1.0$.
Figure~\ref{fig:19}(d,e,f) shows $\langle V_x\rangle$, $\langle V_y\rangle$,
$\langle V\rangle$, and $\theta_{sk}$ versus $f$ for a sample with
 $\alpha_{m}/\alpha_{d} = 3.04$ and $\theta^{\rm int}_{sk} = -71.8^\circ$.
The pinned regions are reduced in width
and occur away from the commensurate fillings. 
There is still directional locking to $\theta_{sk}=-45^\circ$ near $f=1.0$, as
indicated in Fig.~\ref{fig:19}(f) by the $1/1$ label, and
the velocity boosted regime in Fig.~\ref{fig:19}(e) is more extended.

\begin{figure}
\includegraphics[width=3.5in]{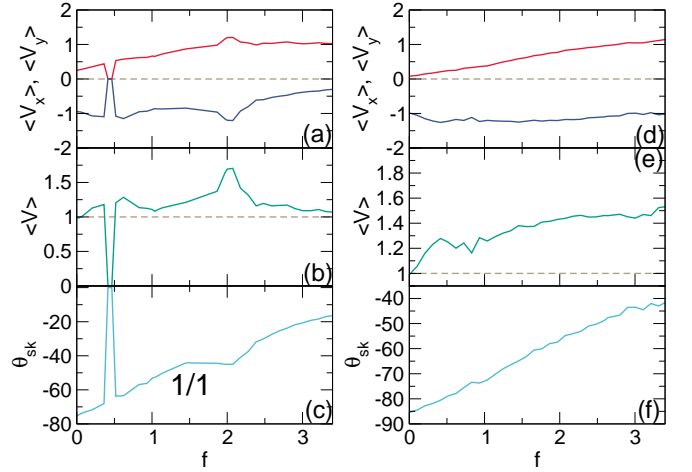}
\caption{
Behavior under varied filling fraction $f$ for a system with a square pinning
array, $F_p=0.75$, and $F_D=1.0$.
(a,d) $\langle V_x\rangle$ (red) and $\langle V_y\rangle$ (blue) versus $f$.
(b,e) $\langle V\rangle$ versus $f$.
(c,f) $\theta_{sk}$ versus $f$.
(a,b,c) $\alpha_m/\alpha_d=4.39$ and $\theta_{sk}^{\rm int}=-77.16^\circ$.
(d,e,f) $\alpha_m/\alpha_d=13.28$ and $\theta_{sk}^{\rm int}=-85.7^\circ$.
}
\label{fig:20}
\end{figure}

In Fig.~\ref{fig:20}(a,b,c) we plot $\langle V_x\rangle$, $\langle V_y\rangle$,
$\langle V\rangle$, and $\theta_{sk}$ versus $f$ for a system with
$\alpha_{m}/\alpha_{d} = 4.39$ and
$\theta^{\rm int}_{sk} = -77.16^\circ$.
There is a single pinned region near $f = 0.5$ 
and a peak in the magnitude of $\langle V_x\rangle$, $\langle V_y\rangle$,
and $\langle V\rangle$
at $f = 2.0$, but there is no 
peak near $f = 1.0$.
The peak
near $f = 2.0$ falls at the end of an extended window of locking
to $-45^\circ$, shown as the $1/1$ step in $\theta_{sk}$ in Fig.~\ref{fig:20}(c).
Over most of the range of $f$ shown,
there is a  strong velocity boost in $\langle V\rangle$, as
indicated by $\langle V\rangle$ running above the $\langle V\rangle=1.0$ line
in Fig.~\ref{fig:20}(b).
Figure~\ref{fig:20}(d,e,f) shows $\langle V_x\rangle$, $\langle V_y\rangle$,
$\langle V\rangle$, and $\theta_{sk}$ for a sample with
$\alpha_{m}/\alpha_{d} = 13.28$
and $\theta^{\rm int}_{sk} = -85.7^\circ$,
where there is no longer a pinned  
phase and only a small $1/1$ locking
step appears near $f = 3.0$.
Here the magnitude of $\theta_{sk}$ 
decreases nearly monotonically with increasing $f$,
while $\langle V\rangle$ shows 
an increasing velocity boost as $f$ becomes larger.
For large Magnus forces such as this, the
dynamics become increasingly disordered, destroying the directional locking.

\begin{figure}
\includegraphics[width=3.5in]{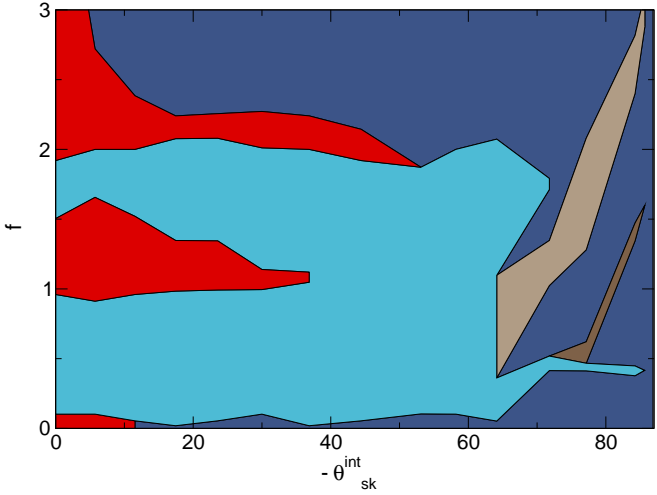}
\caption{
Dynamic phase diagram as a function of filling fraction  $f$ versus
intrinsic skyrmion Hall angle $-\theta_{sk}^{\rm int}$ for the system from
Figs.~\ref{fig:18}, \ref{fig:19}, and \ref{fig:20} with a square pinning array,
$F_p=0.75$, and $F_D=1.0$.
The pinned region is light blue,
the red region is motion with $\theta_{sk}=0.0^\circ$,
the brown region is motion with $\theta_{sk}=-45^\circ$,
the dark brown region is motion with $\theta_{sk}=\arctan(1/2)$,
and the dark blue region is motion with finite $\theta_{sk}$.
Regions with and without velocity boost are not distinguished in
this diagram.
        }
\label{fig:21}
\end{figure}

From the features in the transport curves and
the behavior of  $\theta_{sk}$
for the system in
Figs.~\ref{fig:18}, \ref{fig:19}, and \ref{fig:20},
we construct 
a dynamic phase diagram as a function of $f$ versus $-\theta_{sk}^{\rm int}$
in Fig.~\ref{fig:21}.
We highlight
the pinned phase,
locking to the $x$ direction with $\theta_{sk}=0^\circ$,
locking to $\theta_{sk}=-45^{\circ}$ or the $1/1$ direction,
locking to $\theta_{sk}=\arctan(1/2)$ or the $1/2$ direction,
and motion at a finite skyrmion Hall angle. No distinction is made between
regions with and without a velocity boost in this figure.
For $-\theta^{\rm int}_{sk} < 45^\circ$, there are 
extended regions of locking in the $x$-direction,
while the additional directional
locking effects appear only for higher Magnus fores with
$-\theta^{\rm int}_{sk} > 60^\circ$.
Near $f = 1.0$ and $2.0$, we find extended
regions where $\theta_{sk} = 0^\circ$.
There are extended regions of velocity boosting (not shown) which appear when
$-\theta^{\rm int}_{sk} > 45^\circ$.

\section{Triangular Pinning Arrays}

\begin{figure}
\includegraphics[width=3.5in]{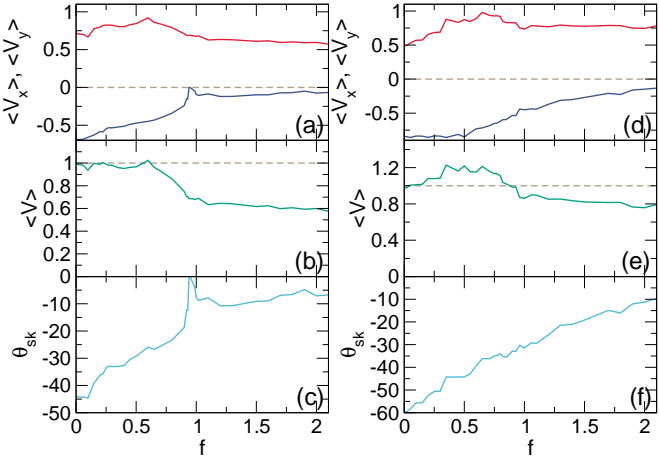}
\caption{
Behavior under varied filling fraction $f$ for a system with a triangular pinning array,
$F_p=0.5$, and $F_D=1.0$.
(a,d) $\langle V_x\rangle$ (red) and $\langle V_y\rangle$ (blue) versus $f$.
(b,e) $\langle V\rangle$ versus $f$.
(c,f) $\theta_{sk}$ versus $f$.
(a,b,c) $\alpha_m/\alpha_d=1.0$ and $\theta_{sk}^{\rm int}=-45^\circ$.
(d,e,f) $\alpha_{m}/\alpha_d = 1.73$ and $\theta_{sk}^{\rm int} = -60^\circ$.
        }
\label{fig:22}
\end{figure}

If the square pinning array is replaced with a triangular pinning array,
similar behavior occurs.
In Fig.~\ref{fig:22}(a,b,c) we plot $\langle V_x\rangle$, $\langle V_y\rangle$,
$\langle V\rangle$, and $\theta_{sk}$ versus $f$ for a sample with
triangular pinning, $F_p=0.5$, $F_D=1.0$,
 $\alpha_{m}/\alpha_{d} = 1.0$, and $\theta^{\rm int}_{sk} = -45.0^\circ$.  
 Here $\theta_{sk}$ goes to zero
 at $f = 1.0$ but not at $f = 2.0$.
 For a triangular pinning array, the background skyrmions
 form a commensurate triangular lattice at $f = 1.0$;
 however, at $f = 2.0$ the skyrmions
 form a honeycomb structure rather than a triangular lattice.
 A similar structure has been observed for superconducting vortices on a 2D
 triangular pinning array
 \cite{Reichhardt98a}.
 Since the honeycomb arrangement is less stable than the triangular arrangement,
 the background skyrmions are not as strongly pinned at $f=2.0$ compared to
 $f=1.0$, and there is less reduction of the drag on the driven skyrmion at $f=2.0$.
 Figure~\ref{fig:22}(d,e,f) shows $\langle V_x\rangle$, $\langle V_y\rangle$,
 $\langle V\rangle$, and $\theta_{sk}$ versus $f$ for the same sample at
 $\alpha_{m}/\alpha_{d} = 1.73$  and $\theta^{\rm int}_{sk} = -60^\circ$,
 where the skyrmion Hall angle does not show any feature at $f = 1.0$. 
 There is a small velocity boost for
 $f < 1.0$, as indicated by the excursion of
 $\langle V\rangle$ above the value
 $\langle V\rangle = 1.0$ in Fig.~\ref{fig:21}(e). 

\begin{figure}
\includegraphics[width=3.5in]{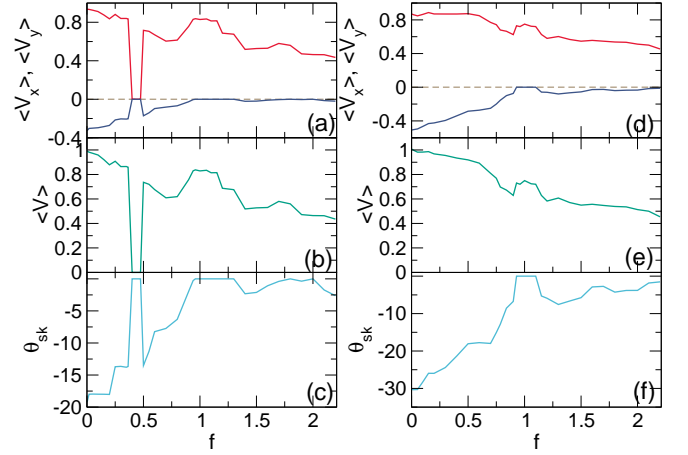}
\caption{
Behavior under varied filling fraction $f$ for a system with a triangular pinning
array, $F_p=0.5$, $F_D=1.0$, and $n_p=0.4882$.
(a,d) $\langle V_x\rangle$ (red) and $\langle V_y\rangle$ (blue) versus $f$.
(b,e) $\langle V\rangle$ versus $f$.
(c,f) $\theta_{sk}$ versus $f$.
(a,b,c) $\alpha_m/\alpha_d=0.374$ and $\theta_{sk}^{\rm int}=-20.5^\circ$.
(d,e,f)
$\alpha_{m}/\alpha_d = 0.658$ and $\theta_{sk}^{\rm int} = -33.4^\circ$.
        }
\label{fig:23}
\end{figure}

\begin{figure}
\includegraphics[width=3.5in]{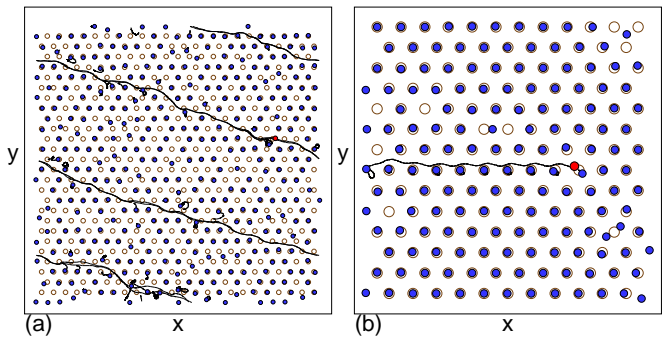}
\caption{
Images of samples containing a triangular pinning lattice showing background skyrmions
(blue filled circles), pinning site locations (brown circles), the driven skyrmion
(red filled circle), and the skyrmion trajectories during a fixed time window for the
system in Fig.~\ref{fig:22}(a,b,c) with $F_p=0.5$, $F_D=1.0$, 
$\alpha_m/\alpha_d=1.0$, and $\theta_{sk}^{\rm int}=-45^\circ$.
(a) $f = 0.6$
(b)
A portion of the sample at $f= 1.0$, where the motion is only along the $x$-direction.  
        }
\label{fig:24}
\end{figure}

In Fig.~\ref{fig:23}(a,b,c) we plot $\langle V_{x}\rangle$, $\langle V_y\rangle$,
$\langle V\rangle$, and $\theta_{sk}$ versus $f$
for a triangular pinning system with
$\alpha_{m}/\alpha_{d} = 0.374$ and $\theta^{\rm int}_{sk} = -20.5^\circ$.
There is a pinned region near $f = 0.4$
along with a region of $\theta_{sk} = 0.0^\circ$ 
centered at $f = 1.0$ that coincides with a peak in
the magnitudes of $\langle V_{x}\rangle$ and $\langle V_{y}\rangle$. 
Figure~\ref{fig:23}(d,e,f) shows the same quantities in a sample with
$\alpha_{m}/\alpha_{d} = 0.658$ and $\theta^{\rm int}_{sk} = -33.4^\circ$,
where a peak appears
in $\langle V_{x}\rangle$ and $\langle V\rangle$ for $f = 1.0$.
We do not see the same commensurate 
effects at $f = 2.0$ due to the honeycomb ordering. 

Figure~\ref{fig:24}(a) illustrates the skyrmion trajectories
for the system in
Fig.~\ref{fig:22}(a,b,c) at $f = 0.6$ where
the skyrmion motion is along
$\theta_{sk} \approx -18^\circ$.
The driven skyrmion interacts both directly with
the  pinning sites and with the background skyrmions, and its
trajectory is disordered.
In Fig.~\ref{fig:23}(b),
the same system at $f = 1.0$
exhibits channeling motion along the $\theta_{sk}=0^\circ$ symmetry
direction of the pinning array.

\section{Discussion}

We expect our
results should be robust for any pinning geometries where commensuration effects
occur.
At matching fillings where the background skyrmions are ordered,
the skyrmion Hall angle will be reduced and there
can be a velocity boost effect.
Other possible pinning array geometries 
include rectangular \cite{Reichhardt01b,Velez02},
1D periodic \cite{LeThien16} or
even quasiperiodic \cite{Kemmler06,Reichhardt11}.
We also expect that thermal effects could enhance some of the
phenomena we observe.
For example, at commensurate conditions the system forms an ordered structure 
that should be more resistant to thermal fluctuations,
while at incommensurate fillings,
the effectiveness of the pinning is reduced so that
creep effects should be more prominent.
In superconducting vortex systems with periodic pinning, commensuration
effects are generally enhanced at higher temperatures.
This effect could also be important when the point particle approximation
breaks down,
such as if multiple skyrmions become trapped at a single pinning site and
experience shape distortions, or if internal skyrmion modes are excited.
We also
expect that this system could exhibit a
variety of phase locking phenomena \cite{Reichhardt15b,Sato20}
if ac driving were introduced.
Such effects should be strongly enhanced near commensurate conditions.
Our results should be general to other systems coupled to 
periodic substrates where gyroscopic effects come into play.  

\section{Summary}

We have numerically examined the dynamics of individual skyrmions
driven through an assembly of other skyrmions in the presence of a
two-dimensional periodic pinning array.
The Magnus force causes the driven skyrmion
to move with a finite skyrmion Hall angle.
Under a constant driving force, we find
a non-monotonic dependence
of the skyrmion Hall angle on the density of the background skyrmions.
In general, the skyrmion
Hall angle drops to zero or is reduced in magnitude
at commensurate conditions when the number of skyrmions is equal to
an integer multiple of the number of pinning sites.
There is also a peak in the net skyrmion 
velocity at the commensurate filling.
At incommensurate fillings, the skyrmion Hall angle
becomes finite again, but there
is generally a decrease in the skyrmion Hall angle with
increasing filling fraction.
At commensurate fillings we find a two step depinning process and
the motion of the driven skyrmion is well ordered,
while at incommensurate fillings there is a single step depinning transition
with disordered motion. 
For larger Magnus forces,
additional locking effects can appear in which
the motion of the driven skyrmion locks to different symmetry directions
of the pinning lattice. 
In some cases, we find that increasing the Magnus force
can enhance the effectiveness of the pinning,
which is opposite to the behavior observed
in a system with random pinning.
This occurs because the Magnus force reduces
the amount of channeling along the $x$ direction.
We show that our results are robust for both square and triangular pinning
arrays.
We also map out dynamical phase diagrams as a function of
varied pinning strength, Magnus force contribution, and filling fraction.

\begin{acknowledgments}
We gratefully acknowledge the support of the U.S. Department of
Energy through the LANL/LDRD program for this work.
This work was supported by the US Department of Energy through
the Los Alamos National Laboratory.  Los Alamos National Laboratory is
operated by Triad National Security, LLC, for the National Nuclear Security
Administration of the U. S. Department of Energy (Contract No. 892333218NCA000001).
\end{acknowledgments}

\bibliography{mybib}

\end{document}